\documentstyle[11pt,aaspp4]{article}
\def\lineunits{ergs\ s$^{-1}$\,cm$^{-2}$}
\def\contunits{ergs\ s$^{-1}$\,cm$^{-2}$\,\AA$^{-1}$}
\def\Fvar{\ifmmode F_{\rm var} \else $F_{\rm var}$\fi}
\def\Rmax{\ifmmode R_{\rm max} \else $R_{\rm max}$\fi}
\def\tcent{\ifmmode \tau_{\rm cent} \else $\tau_{\rm cent}$\fi}
\def\tpeak{\ifmmode \tau_{\rm peak} \else $\tau_{\rm peak}$\fi}

\def\mm{\ifmmode \phantom{$-$} \else \phantom{$-$}\fi}
\def\IUE{{\it IUE}}

\def\arcsecpoint{$''\!.$}

\def\kms{\ifmmode {\rm km\ s}^{-1} \else km s$^{-1}$\fi}
\def\Msun{\ifmmode M_{\odot} \else $M_{\odot}$\fi}
\def\Lsun{\ifmmode L_{\odot} \else $L_{\odot}$\fi}
\def\qo{\ifmmode q_{\rm o} \else $q_{\rm o}$\fi}
\def\Ho{\ifmmode H_{\rm o} \else $H_{\rm o}$\fi}
\def\ho{\ifmmode h_{\rm o} \else $h_{\rm o}$\fi}

\def\vFWHM{\ifmmode v_{\mbox{\tiny FWHM}} \else
            $v_{\mbox{\tiny FWHM}}$\fi}
\def\CCF{\ifmmode F_{\it CCF} \else $F_{\it CCF}$\fi}
\def\ACF{\ifmmode F_{\it ACF} \else $F_{\it ACF}$\fi}
\def\Halpha{\ifmmode {\rm H}\alpha \else H$\alpha$\fi}
\def\Hbeta{\ifmmode {\rm H}\beta \else H$\beta$\fi}
\def\Hgamma{\ifmmode {\rm H}\gamma \else H$\gamma$\fi}
\def\Hdelta{\ifmmode {\rm H}\delta \else H$\delta$\fi}
\def\Lya{\ifmmode {\rm Ly}\alpha \else Ly$\alpha$\fi}
\def\Lyb{\ifmmode {\rm Ly}\beta \else Ly$\beta$\fi}

\def\ciii{\ifmmode {\rm C}\,{\sc iii} \else C\,{\sc iii}\fi}
\def\civ{\ifmmode {\rm C}\,{\sc iv} \else C\,{\sc iv}\fi}

\def\oiii{O\,{\sc iii}}
\def\o5007{[O\,{\sc iii}]\,$\lambda5007$}

\def\feii{Fe\,{\sc ii}}

\lefthead{Peterson et al.}
\righthead{Long-Term Monitoring of NGC 5548}

\begin{document}
\title{Steps Toward Determination of the Size and Structure
of the Broad-Line Region in Active Galactic Nuclei.\ XVI.\
A Thirteen-Year Study of Spectral Variability in NGC~5548}

\author{
B.\,M.~Peterson,\altaffilmark{1}
P.~Berlind,\altaffilmark{2}
R.~Bertram,\altaffilmark{1}$^{,}$\altaffilmark{3}
K.~Bischoff,\altaffilmark{4}
N.\,G.~Bochkarev,\altaffilmark{5}
N.~Borisov,\altaffilmark{6} 
A.\,N.~Burenkov,\altaffilmark{6}$^{,}$\altaffilmark{7}
M.~Calkins,\altaffilmark{2}
L.~Carrasco,\altaffilmark{8}
V.\,H.~Chavushyan,\altaffilmark{8} 
R.~Chornock,\altaffilmark{9}
M.~Dietrich,\altaffilmark{10}
V.\,T.~Doroshenko,\altaffilmark{11}
O.\,V.~Ezhkova,\altaffilmark{12}
A.\,V.~Filippenko,\altaffilmark{9}
A.\,M.~Gilbert,\altaffilmark{9}
J.\,P.~Huchra,\altaffilmark{2}
W.~Kollatschny,\altaffilmark{4}
D.\,C.~Leonard,\altaffilmark{9}$^{,}$\altaffilmark{13}
W.~Li,\altaffilmark{9}
V.\,M.~Lyuty,\altaffilmark{11}
Yu.\,F.~Malkov,\altaffilmark{14}
T.~Matheson,\altaffilmark{9}$^{,}$\altaffilmark{2}
N.\,I.~Merkulova,\altaffilmark{14}$^{,}$\altaffilmark{7}
V.\,P~Mikhailov,\altaffilmark{6}
M.~Modjaz,\altaffilmark{9}$^{,}$\altaffilmark{2}
C.\,A.~Onken,\altaffilmark{1}
R.\,W.~Pogge,\altaffilmark{1}
V.\,I.~Pronik,\altaffilmark{14}$^{,}$\altaffilmark{7}
B.~Qian,\altaffilmark{15}
P.~Romano,\altaffilmark{1}
S.\,G.~Sergeev,\altaffilmark{14}$^{,}$\altaffilmark{7}
E.\,A.~Sergeeva,\altaffilmark{14}$^{,}$\altaffilmark{7}
A.\,I.~Shapovalova,\altaffilmark{6}$^,$\altaffilmark{7}
O.\,I.~Spiridonova,\altaffilmark{6}
J.~Tao,\altaffilmark{15}
S.~Tokarz,\altaffilmark{2}
J.\,R.~Valdes,\altaffilmark{8}
V.\,V.~Vlasiuk,\altaffilmark{6}
R.\,M.~Wagner,\altaffilmark{1}$^{,}$\altaffilmark{3}
and B.\,J.~Wilkes\altaffilmark{2}
}
\altaffiltext{1}
        {Department of Astronomy, The Ohio State University,
		140 West 18th Avenue, Columbus, OH  43210--1173.
	peterson, onken, pogge, promano@astronomy.ohio-state.edu}
\altaffiltext{2}
	{Harvard-Smithsonian Center for Astrophysics,
	60 Garden Street, Cambridge, MA 02138.
	pberlind, huchra, tmatheson, mmodjaz, belinda@cfa.harvard.edu}
\altaffiltext{3}
	{Mailing address: Steward Observatory, University of Arizona,
	Tucson, AZ  85721.
	rayb, rmw@as.arizona.edu}
\altaffiltext{4}
	{Universit\"{a}ts-Sternwarte G\"{o}ttingen, Geismarlandstr.\ 11,
	D--37083 G\"{o}ttingen, Germany.
	bischoff@uni-sw.gwdg.de, wkollat@gwdg.de}
\altaffiltext{5}
	{Sternberg Astronomical Institute, Lomonosov Moscow
	State University, Universitetskij Prosp.\ 13, Moscow 119992, 
Russia.
	boch@sai.msu.ru}
\altaffiltext{6}
	{Special Astrophysical Observatory, Russian Academy of Sciences,
	Nizhnij Arkhyz, Karachai-Cherkess Republic, 357147, Russia.
	ban, mvp, ospir, ashap, ospir, vvlas@sao.ru}
\altaffiltext{7} 
	 {Also: Isaac Newton Institute of Chile, Special Astrophysical
	Observatory Branch and Crimean Branch.}
\altaffiltext{8} 
	{Instituto Nacional de Astrofisica,
	Optica y Electronica, Apartado Postal 51, CP 72000, 
	Puebla, Mexico. carrasco, vahram, jvaldes@inaoep.mx}
\altaffiltext{9} 
	{\raggedright Department of Astronomy, 
	University of California, Berkeley, CA 94720--3411.
	rchornock, alex, agilbert, wli@astro.berkeley.edu}
\altaffiltext{10}
	{Department of Astronomy, University of Florida,
	211 Bryant Space Science Center, Gainesville, FL 32611--2055.
	dietrich@astro.ufl.edu}
\altaffiltext{11}
	{Crimean Laboratory of the Sternberg Astronomical Institute, 
	p/o Nauchny, Crimea, 98409, Ukraine.
	doroshen, lyuty@sai.crimea.ua}
\altaffiltext{12}
	{Ulugbek Astronomical Institute, Academy of Science of 
Uzbekistan,
	Astronomicheskaya ul. 33, Tashkent, 700052, Uzbekistan.
	oezh@sai.msu.ru}
\altaffiltext{13}
	{\raggedright Present address: Department of Astronomy, 
University of
	Massachusetts, Amherst, MA 01003-9305.
	 leonard@nova.astro.umass.edu}
\altaffiltext{14}
	{\raggedright Crimean Astrophysical Observatory, 
	P/O Nauchny, 98409 Crimea, Ukraine. 
	nelly@astro.crao.crimea.ua, vpronik, sergeev, 
selena@crao.crimea.ua}
\altaffiltext{15}
	{Shanghai Astronomical Observatory, Shanghai 200030,
	China.
	taojun@center.shao.ac.cn}

\clearpage

\begin{abstract}
We present the final installment of an intensive 13-year study
of variations of the optical continuum and broad \Hbeta\ emission line
in the Seyfert 1 galaxy NGC~5548. The data base consists of
1530 optical continuum measurements and 1248 \Hbeta\ measurements.
The \Hbeta\ variations follow the continuum variations closely,
with a typical time delay of about 20 days. However, a year-by-year
analysis shows that the magnitude of emission-line time delay is 
correlated
with the mean continuum flux. We argue that the data are
consistent with the simple model prediction
between the size of the broad-line region and the ionizing luminosity,
$r \propto L_{\rm ion}^{1/2}$. Moreover, the 
apparently linear nature of the correlation between
the \Hbeta\ response time and the nonstellar optical continuum $F_{\rm 
opt}$
arises as a consequence of the changing shape of the
continuum as it varies, specifically
$F_{\rm opt} \propto F_{\rm UV}^{0.56}$.
\end{abstract}

\setcounter{footnote}{0}

\section{Introduction}

The nature of the broad-line region (BLR) 
is arguably one of the major remaining mysteries
in active galactic nuclei (AGNs). Whereas the
supermassive black hole/accretion-disk paradigm
has become increasingly more secure, there is still
no consensus about the origin of the broad emission
lines that are prominent features of the UV/optical
spectra of these sources. This is not to say that
progress has not been made. Years of spectroscopic
study have uncovered a rich phenomenology
(see Sulentic, Marziani, \& Dultzin-Hacyan 2000 for a recent review).

Of particular importance has been the recognition
that the emission-line fluxes vary in response
to continuum variations, with a small time
delay (days to weeks for Seyfert galaxies)
due to light travel-time effects within the BLR.
Well before these time delays were first accurately
measured, this led to the seminal paper on
``reverberation mapping'' (Blandford \& McKee 1982),
a tomographic method of determining the 
structure and kinematics of the BLR. While the
full potential of reverberation mapping has yet to be
realized, use of this technique has provided 
BLR sizes in approximately three dozen AGNs
(see compilations by Wandel, Peterson, \& Malkan 1999
and Kaspi et al.\ 2000). Even more importantly,
measurement of the emission-line time delays (or ``lags''),
combined with measurements of the line width, have
led to estimates of the masses of the central black holes.

In late 1988, we began a program of spectroscopic monitoring of
optical variations in the Seyfert 1 galaxy NGC~5548.
This activity was organized by an informal consortium
called the ``International AGN Watch'' (Alloin et al.\ 1994;
Peterson 1999). 
Initially, the optical monitoring program was undertaken in support of
an ultraviolet monitoring campaign carried out with the
{\it International Ultraviolet Explorer} (\IUE\,; Clavel et al.\ 1991
[the first paper in the same series
as this work, and hereafter referred to as Paper I]).
The resulting optical data set (Peterson et al.\ 1991, hereafter Paper 
II) 
proved
to be so rich that it was decided to continue this program.
Since that time, a number of additional contributions to this series
(Peterson et al.\ 1992 [Paper III];
Peterson et al.\ 1994 [Paper VII];
Korista et al.\ 1995 [Paper VIII]; and
Peterson et al.\ 1999 [Paper XV])
have described our continuing optical monitoring program.
A complete summary of the work on NGC~5548 by this consortium
appears in Table 1. Reviews of the progress of this program 
are also available (e.g., Peterson 1993, 2001; Netzer \& Peterson 
1997).

This paper represents the
final installment of our 13-year program of optical monitoring
of NGC~5548. As is customary for the papers in this series, we will
focus primarily on the data and the salient results of
cross-correlation analysis of the continuum and emission-line
light curves. Further in-depth analysis will be left to subsequent 
papers.
In \S\,2, we describe the observations, data reduction, and
intercalibration procedures that we have used to
construct a homogeneous data base of optical continuum
and \Hbeta\ emission-line fluxes. In \S\,3, we describe
the time-series analysis that we have undertaken
to determine the time scale for response of \Hbeta\
to continuum variations. In \S\,4, we discuss some
of the implications of our study.
Our results are summarized in \S\,5.
We also note that some of the data
presented here have already appeared in other contexts,
specifically as part of a short but intensive monitoring
program carried out in 1998 June (Dietrich et al.\ 2001)
and as part of a long-term photometric monitoring campaign
(Doroshenko et al.\ 2001).

\section{Observations and Data Analysis}

\subsection{Spectroscopic Observations}

Here we analyze spectroscopic observations made
between 1995 November 22 (Julian Date = JD2450044) and
2001 September 21 (JD2452174); UT dates are used throughout this paper.
During this six-year period, a total of 543 individual
spectra were obtained, as summarized 
in Table 2. Column (1) gives a code for each data set; these
are the same codes that have been used throughout the 
NGC 5548 monitoring project.
A more complete log of observations, as has been
published in our previous papers on NGC 5548, can be found
at the International AGN Watch 
site on the World-Wide Web\footnote{The light curves 
and complete logs of observation are
available in tabular form at URL 
{\sf http://www.astronomy.ohio-state.edu/$\sim$agnwatch/}.
All publicly available International AGN Watch data can be accessed
at this site, which also includes complete references to
published AGN Watch papers.}.

The data obtained during Year 8 of this program 
(see Table 1) were previously
published in Paper XV. Here we completely reanalyze the
data from Year 8 because 46 additional spectra from set ``F''
have become available to us since Paper XV was 
published\footnote{We also note in passing that we have
included in the AGN Watch archive
five additional set ``H'' spectra from Years 3 and 4
(details are provided on the AGN Watch website),
although these data have not been incorporated into
the light curves that are currently available.}.

The spectroscopic images were processed by individual observers
in standard
fashion for CCD frames, including bias subtraction, 
flat-field correction, wavelength calibration,
and flux calibration based on standard-star observations.
The flux calibration is then refined 
by scaling each spectrum to a constant value of the
\o5007 narrow-line flux, as we have done in each
previous paper in this series. All spectra are
gray-scaled in flux (i.e., adjusted by a constant
multiplicative factor) to a constant flux of
$F(\mbox{\o5007}) = 5.58 \times 10^{-13}$\,\lineunits.
This absolute flux was determined from spectra made under photometric
conditions during the first year of this program 
(Paper II). Scaling is accomplished automatically
by use of the van Groningen \& Wanders (1992)
spectral scaling software, as described in Paper XV.

After flux scaling, measurements of each of the spectra are made.
The continuum flux at $\sim5100$\,\AA\ (in the rest frame of
NGC 5548, $z = 0.0167$) is determined by averaging the flux in 
the 5185--5195\,\AA\ bandpass (in the observed frame).
The \Hbeta\ emission-line flux is measured by assuming a linear
underlying continuum between $\sim4790$\,\AA\ and $\sim5170$\,\AA,
and integrating the flux above this continuum between 4795\,\AA\ and
5018\,\AA\ (all wavelengths in the observed frame). The long-wavelength
cutoff of this integration band misses some of the \Hbeta\ flux
underneath [\oiii]\,$\lambda4959$, but avoids the need to
estimate the \feii\ contribution to this feature and still gives
a good representation of the \Hbeta\ variability. We also note
that no attempt has been made to correct for contamination of
the line measurement by the {\em narrow-line}\ component of 
\Hbeta, which is of course expected to be constant.

Even after scaling all of the spectra to a common
value of the \o5007\ flux, there are systematic differences
between the light curves produced from data
obtained at different telescopes. 
As in our previous papers,
we correct for the small offsets between the light curves from
different sources in a simple, but effective, fashion. We
attribute these small relative offsets
to aperture effects (Peterson et al.\ 1995), although the
procedure we use also corrects for other unidentified systematic
differences between data sets.  We
define a point-source correction factor $\varphi$ by the equation
\begin{equation}
\label{eq:defphi}
F(\Hbeta)_{\rm true} = \varphi F(\Hbeta)_{\rm observed}.
\end{equation}
This factor accounts for the fact that different apertures
result in different amounts of light loss for the 
point-spread function (which describes the surface-brightness
distribution of both the broad lines and the AGN continuum
source) and the partially extended narrow-line region.

After correcting for aperture effects on the point-spread function
to narrow-line ratio, we also correct
for the different amounts of starlight admitted by
different apertures. An extended source correction $G$ 
is thus defined as
\begin{equation}
\label{eq:defG}
F_{\lambda}(5100\,{\textstyle {\rm \AA}})_{\rm true} = \varphi 
F_{\lambda}(5100\,{\textstyle {\rm \AA}})_{\rm observed} - G.
\end{equation}

Intercalibration of the individual data
sets is then accomplished  by
comparing pairs of nearly simultaneous observations from different
data sets to determine 
for each data set the values of the constants $\varphi$ and $G$
that are needed to adjust the emission-line and continuum
fluxes to a common scale. Furthermore, the formal uncertainties
in $\varphi$ and $G$ reflect the uncertainties in the 
individual data sets, so we can determine the nominal
uncertainties for each data set if we assume that the errors add
in quadrature. 

As in our previous work, 
the data are adjusted  relative to data set ``A'' 
because these data are fairly extensive, 
overlap well with most of the other data sets, and were obtained 
through
a reasonably large aperture (5\arcsecpoint0 $\times$ 7\arcsecpoint5).
As in Paper XV, fractional uncertainties 
of $\sigma_{\rm cont}/F_{\lambda}(5100$\,\AA) $\approx 0.020$ and
$\sigma_{\rm line}/F(\Hbeta) \approx 0.020$ for the continuum
and \Hbeta\ line, respectively, are adopted for the similar, 
large-aperture, high-quality data sets ``A'' and ``H,''
based on the differences between closely spaced observations
within these sets (see also Peterson et al.\ 1998a). 
For the other data sets, it was possible to
estimate the mean uncertainties in the measurements by
comparing them to measurements from other sets for
which the uncertainties are known and by assuming that
the uncertainties for each set add in quadrature.

The intercalibration constants we use for each data set are
given in Table 3, and these constants are used with 
equations (\ref{eq:defphi}) and (\ref{eq:defG}) to adjust
the spectral measurements. We note that there were a few cases in
which the intercalibration constants had to be
evaluated on a year-to-year basis
to effect an acceptable intercalibration with the 
rest of the data. 

\subsection{Photometric Observations}

In addition to the spectroscopic observations,
numerous $V$-band photometric observations of
NGC~5548 were made. The sources of these
observations are summarized in Table 4
and a more complete log of observations is
available on the AGN Watch website.
Many of the observations used in this analysis were also
included in a study by Doroshenko et al.\ (2001).

In order to effect an intercalibration with the
spectroscopic light curve, we compared closely
spaced pairs of spectroscopic and photometric
observations and performed a least-squares fit
to the equation
\begin{equation}
\label{eq:Vfit}
F_{\lambda}(5100\,{\textstyle {\rm \AA}})
= a + b F_V,
\end{equation}
where the magnitude $V \propto -2.5\log F_V$.
This fit was based on pairs of observations obtained
within one day of each other.
The raw $V$-band fluxes were adjusted using eq.\ (\ref{eq:Vfit})
to the $F_{\lambda}(5100\,{\textstyle {\rm \AA}})$ scale defined
by the spectrophotometry.

\subsection{The Merged Light Curves}

A complete continuum light curve was assembled by merging
the spectrophotometric and $V$-band photometric light curves
described above.
The final light curves for the optical continuum and \Hbeta\ fluxes
were then constructed by computing a weighted average of points
separated in time by less than 0.1 day. The final light
curves for Years 8 through 13 are given in Table 5.
It is important to
note that all measurements are in the observer's
frame and are uncorrected for Galactic extinction.
Also, the continuum flux includes a contribution from
the host galaxy, which we estimate to be
$3.4 \times 10^{-15}$\,\contunits\ (Romanishin et al.\ 1995),
and the emission-line flux includes the narrow-line
component, which we estimate to be
$\sim 8.2\times 10^{-14}$\,\lineunits\ (Wanders \& Peterson 1996);
both of these values are in the observer's reference frame. 

\section{Variability Analysis}

The data given in Table 5 can be combined with our previously
published light curves (Papers II, III, VII, VIII, and XV), 
yielding homogeneous light
curves that cover a span of 4756 days. The combined data are shown in 
Figure 1.
These light curves are comprised of 1530 continuum measurements
and 1248 line measurements. 
In this section, we will summarize the basic characteristics of
the thirteen-year homogeneous data base.

\subsection{Characteristics of the Data Base}

In Tables 6 and 7, we provide a summary of the basic
characteristics of the continuum and \Hbeta\ 
emission-line light curves, respectively, as shown in Figure 1.
Column (1) indicates the subset considered, with time ranges
as defined in Table 1.  The number of observations in each
subset is given in column (2), and columns (3) and (4) give
the average and median intervals, 
respectively, between successive observations. The mean and 
root-mean-square (rms) fluxes, $F_{\lambda}$(5100\,\AA)
in Table 6 and $F$(\Hbeta) in Table 7, are shown in column (5).
Two standard measures of variability, \Fvar\ and \Rmax, are given
in columns (6) and (7), respectively. 
The parameter \Fvar\ is the rms fractional variability,
corrected for measurement error, as defined by 
Rodr\'{\i}guez-Pascual et al.\ (1997), and \Rmax\ is simply
the ratio of maximum to minimum flux. Both of these parameters
are affected by contamination of the measured quantities by
constant-flux components, namely the underlying host galaxy in
the case of the continuum, and the narrow \Hbeta\ emission
component in the case of the line. If we use the values for the host-
galaxy
contribution and the narrow-line \Hbeta\ flux given in the
last section, the values of \Fvar\ increase
to 0.380 and 0.242 for the continuum and \Hbeta\ line, respectively,
and the values of \Rmax\ increase to $9.44\pm1.41$ and
$6.26\pm0.62$, respectively. It is notable that
the optical 
nonstellar continuum in NGC 5548 has varied by an order of magnitude
over this 13-year monitoring program.

\subsection{Time-Series Analysis}

As in Paper XV and previous papers, we compute the time delay between
continuum variations and the \Hbeta\ response by cross-correlation
of the  continuum and emission-line
light curves. We perform the cross-correlation analysis in two
ways, using the interpolation cross-correlation function (ICCF) method
of Gaskell \& Sparke (1986) and Gaskell \& Peterson (1987), and the
discrete correlation function (DCF) method of Edelson \& Krolik (1988).
In both cases, we use the implementation described by White \& Peterson
(1994). 

The results of the cross-correlation analysis are shown in
Figure 2 and summarized in Table 8. The entries in Table 8
give the subset used in the analysis (column 1),
the centroid \tcent\ of the ICCF (column 2),
the peak \tpeak\ of the ICCF (column 3),
and $r_{\rm max}$, the value of the ICCF at \tpeak.
The value of \tcent\ is computed using only those points
near the principal peak with values exceeding 
0.8$r_{\rm max}$.
The uncertainties quoted for \tcent\ and \tpeak\
have been computed using the model-independent
FR/RSS Monte-Carlo method described by Peterson et al.\ (1998b).

\section{Discussion}

It was noted in Paper XV that statistically significant
year-to-year changes in
the \Hbeta\ lag have occurred, and that 
these changes are correlated with the mean continuum
flux. The additional data described here confirm this
result and allow us to investigate it further.
In Figure 4,  we plot \tcent\ as a function of
the mean starlight-corrected
continuum flux for each year
$\bar{F}_{\rm opt} = \langle F_{\lambda}({\rm 5100\,\AA})\rangle - 
F_{\rm gal}$, where 
$F_{\rm gal}$ is our best estimate of the
host-galaxy starlight contribution through our standard
aperture, $3.4 \times 10^{-15}$\,\contunits\ at
5100\,\AA\ (Romanishin et al.\ 1995); $\bar{F}_{\rm opt}$ is
thus the mean nonstellar flux from the AGN component only.
We add to this a measurement from an earlier monitoring
campaign at Wise Observatory (Netzer et al.\ 1990),
adjusted as described in Paper XV.
The best-fit power-law relationship between the continuum
flux and time lag is $\tcent \propto \bar{F}_{\rm opt}^{0.95}$, nearly
a linear relationship.

This result can be compared with a simple theoretical
prediction.
Photoionization equilibrium models for the BLR are
characterized by 
(a) the shape of the ionizing continuum incident upon
the BLR gas, and (b) an ionization parameter $U$, which is the ratio
of ionizing photon density to particle density at the
cloud inner face,
\begin{equation}
U = \frac{Q{\rm (H)}}{4 \pi r^2 n_{\rm H} c},
\end{equation}
where $Q{\rm (H)}$ is the rate at which ionizing photons
are produced by the continuum source, $r$ is the separation
between the BLR gas and the continuum source, and 
$n_{\rm H}$ is the BLR particle density. Since the
luminosity in ionizing photons $L_{\rm ion}$ is
proportional to $Q{\rm (H)}$, we expect
that $r^2 \propto L_{\rm ion}/U n_{\rm H}$. 
Essentially, as $L_{\rm ion}$ varies, we expect
that the response in a particular emission line will
be greatest at some particular value of 
the product $U n_{\rm H}$, which will thus lead
to $r \propto L_{\rm ion}^{1/2}$ for any given line;
in other words, the response of a particular line to 
a continuum variation ought to be dominated by
gas in which the change in emissivity is greatest.
In Figure 3, we show a test of this naive prediction
based on  fitting a $r \propto L^{1/2}$ 
(i.e., $\tcent \propto \bar{F}_{\rm opt}^{1/2}$) power law to these 
data. 
The resulting fit is quite poor.

The simple model above makes the implicit assumption that
the continuum does not change shape as it varies. This allows us to
use $L_{\rm opt}$ as a surrogate for $L_{\rm ion}$,
which cannot be measured directly. However, a
flux at an ultraviolet wavelength
closer to the Lyman edge at 912\,\AA\ would obviously 
provide a better surrogate for $L_{\rm ion}$ than $L_{\rm opt}$, and 
such
data do exist in archival form, mostly from our
previous UV monitoring experiments (Papers I and VIII)
on NGC 5548. We have therefore recovered the NEWSIPS-extracted \IUE\
spectra of NGC~5548 obtained since late 1989 from
the {\em IUE Final Archive} in order to effect a direct
comparison of the UV and optical amplitude of continuum variability.
We measured the UV continuum flux $F_{\rm UV} 
\equiv F_{\lambda}(\mbox{1350\,\AA)}$
by averaging the flux in the observed wavelength range 1370--1380\,\AA\ 
for all \IUE\ SWP spectra obtained within one day of 
optical continuum measurements. All optical fluxes within one day
of a UV continuum measurement were used to form a weighted average
optical flux. This procedure yielded 83 pairs of UV/optical continuum
measurements. We then fitted a power-law function
$F_{\rm opt} \propto F_{\rm UV}^{\alpha}$
to these data, as shown in Figure 4, yielding a best-fit
slope $\alpha = 0.56$. Combining this with the above relationship
between \tcent\ and the optical continuum yields
$\tcent \propto F_{\rm UV}^{0.53}$, which is consistent with
the simple model. We note that there is no statistically
significant time delay between the UV and optical continuum
variations (Peterson et al.\ 1998b, and previous papers in this
series).

Thus, the apparently linear correlation between the variations
in the optical continuum and those in the \Hbeta\ emission line
is attributable to similar relationships between the UV and
optical continuum variations on the one hand and the UV continuum
and \Hbeta\ emission-line variations on the other.
This probably accounts for the apparent absence of scatter
(relative to, say, \Lya\ or \civ\,$\lambda1549$) in comparisons
of the optical continuum and \Hbeta\ fluxes of AGNs (e.g., Yee 1980;
Peterson 1997).

It is interesting that our observational result agrees with
the naive theory once we account for the change in the shape
of the continuum between the UV and optical, but that naive
theory supposes that the shape of the ionizing continuum remains
unchanged. Either the shape of the continuum between the
UV and extreme ultraviolet still remains constant, or
the \Hbeta\ line is surprisingly insensitive to the shape of
the continuum.

It should be pointed out that our result differs from the
BLR radius--luminosity relationship found by Kaspi et al.\ (2000),
namely $r \propto L^{0.7}$. The Kaspi et al.\ result describes
how the BLR radius varies from object-to-object as a function of
the mean optical luminosity of the source. 
The relationship discussed here  describes how the BLR radius
in an individual object changes as the luminosity of
the central source varies with time.

\section{Summary}

We have presented an additional five years (1997--2001)
of optical spectroscopic 
and photometric observations of the continuum and \Hbeta\ emission-line
variations in the Seyfert 1 galaxy NGC 5548. We have also
added significantly to the previously published data from 1996,
warranting a complete recalibration of the data for that particular 
year.
The new data expand our temporal coverage of variations
in this AGN to a total of 1530 continuum and 1248 \Hbeta\
measurements obtained over a 4756-day span, beginning in 1988 December 
and terminating in 2001 December. During this period,
the nonstellar optical continuum in NGC~5548 varied by approximately
an order of magnitude in flux.

Analysis of the time delay between continuum variations
and \Hbeta\ response allows us to confirm our earlier finding
(Paper XV)
that the \Hbeta\ lag varies with the mean continuum flux, ranging
from a low value of $\sim 6$ days in 2000 to a high value of
$\sim 26$ days in 1998--1999. We find consistency with
the simple photoionization equilibrium prediction 
$\tcent \propto L_{\rm UV}^{1/2}$.
\bigskip

We are grateful to the Directors and Telescope Allocation
Committees of our various observatories for their support of
this project. Individual investigators have benefited from
the support from a number of grants, including the following:
National Science Foundation grants AST--9420080 (Ohio State University)
and AST--9987938 (University of California, Berkeley);
NASA grants NAG5-8397 (Ohio State University) and
NAG5-3234 (University of Florida); 
US Civilian Research and Development Foundation Award No.\ UP1--2116
(Crimean Astrophysical Observatory and Ohio State University);
Russian Basic Research Foundation Grants  N97-02-17625 and
N00-02-16272a (Sternberg Astronomical Institute and
Special Astrophysical Observatory);
INTAS grant N96-0328 and CONACyT research grants
G28586-E, 28499-E, and 32106-E (INAOE), and
Sonderforschungsbereich grants SFB\,328\,D and SFB\,439.
We thank the referee, Dr.\ M.\ Goad, for suggestions that
led to an improved presentation.



\clearpage

\begin{figure}
\caption{The optical continuum (upper panel) and
\Hbeta\ (lower panel) light curves from 1989 December
to 2001 December. The data are comprised of 1530 continuum
measurements and 1248 emission-line measurements.
The continuum fluxes
are in units of $10^{-15}$\,\contunits, and the line fluxes
are in units of $10^{-13}$\,\lineunits. The horizontal
dashed line is an estimate of the continuum
contribution from the host galaxy (Romanishin et al.\ 1995)
through the standard aperture used here
(5\arcsecpoint0 $\times$ 7\arcsecpoint5). Flux measurements are
in the observer's reference frame and are uncorrected
for Galactic extinction.}
\end{figure}


\begin{figure}
\caption{Cross-correlation functions for the continuum and
\Hbeta\ emission-line light curves of NGC 5548. The various
panels represent subsets of the new data presented here.
The solid line shows the ICCF, and the vertical lines
represent the $1\sigma$ uncertainties associated with the
DCF values. The peaks and centroids of the ICCFs are
summarized in Table 8.}
\end{figure}

\begin{figure}
\caption{The cross-correlation centroid
\tcent\ as a function the yearly mean optical continuum
flux $\langle F_{\lambda}\mbox{(5100\,\AA)} \rangle$,
based on eight years of International AGN Watch
data (filled circles) plus 1988 data from Wise Observatory
(open circle). The vertical dashed line indicates our
best estimate of the host galaxy flux $F_{\rm gal}$
through our standard aperture. 
The solid line is the best-fit power law to
the relationship $\tcent \propto F_{\rm opt}^{\alpha}$, 
where $F_{\rm opt} = F_{\lambda}\mbox{(5100\,\AA)} -
F_{\rm gal}$, i.e., the nonstellar optical continuum flux.
We find a best-fit slope $\alpha = 0.95$. The dotted line
is the best fit to the naive prediction
$\tcent \propto F_{\rm opt}^{1/2}$.}
\end{figure}

\begin{figure}
\caption{The relationship between UV continuum fluxes
$F_{\rm UV}$
measured from NEWSIPS-extracted \IUE\ spectra
and the nonstellar optical continuum $F_{\rm opt}$
for pairs of UV and optical continuum measurements
made within one day of each other.
Shown as a dashed line is the best-fit power law
to these data, $F_{\rm opt} \propto F_{\rm UV}^{0.56}$.
Combining this relationship with that shown in
Figure 4 yields
$\tcent \propto F_{\rm UV}^{0.53}$, consistent with
the simple model prediction 
$\tcent \propto F_{\rm UV}^{1/2}$.}
\end{figure}

\clearpage

%
%
\begin{deluxetable}{lll}
\tablewidth{0pt}
\tablecaption{Summary of Program on NGC 5548}
\tablehead{
\colhead{Year} &
\colhead{Dates} &
\colhead{Comments} \\
\colhead{(1)} &
\colhead{(2)} &
\colhead{(3)} 
}
\startdata
1 & 1988 Dec 14 -- 1989 Aug 7 & 1: UV continuum and lines \\
  & 1988 Dec 14 -- 1989 Oct 10& 2: Optical continuum and H$\beta$ \\
  &                            & 3: Multiple optical lines \\
  &                           & 4: Balmer continuum and Fe\,{\sc ii} 
\\
  &                           & 5: Optical imaging/photometry \\
  &			      & 6: Revised optical continuum \\
2 & 1989 Dec 1 -- 1990 Oct 15 & 6: Optical continuum and H$\beta$ \\ 
3 & 1990 Nov 29 -- 1991 Oct 5 & 7: Optical continuum and H$\beta$ \\ 
4 & 1992 Jan 1 -- 1992 Oct 3 &  7: Optical continuum and H$\beta$ \\ 
5 & 1993 Mar 14 -- 1993 May 27 & 8: UV continuum and lines \\
  & 1992 Nov 27 -- 1993 Sep 25 & 8: Optical continuum and H$\beta$ \\ 
  & 1993 Mar 10 -- 1993 May 14 & 9: EUV continuum \\
6 & 1993 Nov 17 -- 1994 Oct 11 & 10: Optical continuum and H$\beta$ \\ 
7 & 1994 Nov 22 -- 1995 Oct 17 & 10: Optical continuum and H$\beta$ \\ 
8 & 1995 Nov 22 -- 1996 Oct 16 & 10: Optical continuum and H$\beta$ \\ 
  &                           & 11: Revised optical continuum and H$\beta$ \\ 
9 & 1996 Dec 17 -- 1997 Oct 7 & 11: Optical continuum and H$\beta$ \\ 
10& 1997 Nov 22 -- 1998 Sep 28& 11: Optical continuum and H$\beta$ \\ 
11& 1998 Nov 24 -- 1999 Oct 4 &11: Optical continuum and H$\beta$ \\ 
12& 1999 Dec 5 -- 2000 Sep 3 &11: Optical continuum and H$\beta$ \\ 
13& 2000 Nov 30 -- 2001 Dec 21 & 11: Optical continuum and H$\beta$ \\ 
\hline
\multicolumn{3}{l}{References:} \\
1 & Clavel et al.\ 1991 (Paper I) \\
2 & Peterson et al.\ 1991 (Paper II)  \\
3 & Dietrich et al.\ 1993. (Paper IV) \\
4 & Maoz et al.\ 1993 \\
5 & Romanishin et al.\ 1995 \\
6 & Peterson et al.\ 1992 (Paper III) \\
7 & Peterson et al.\ 1994 (Paper VII) \\
8 & Korista et al.\ 1995 (Paper VIII) \\
9 & Marshall et al.\ 1997 \\
10& Peterson et al.\ 1999 (Paper XV) \\
11& This work.
\enddata

\end{deluxetable}
%
%
\begin{deluxetable}{llc}
\tablewidth{0pt}
\tablecaption{Sources of Spectroscopic Observations, Years 8--13}
\tablehead{
\colhead{Data} &
\colhead{ } &
\colhead{Number of} \\
\colhead{Set} &
\colhead{Telescope and Instrument} &
\colhead{Spectra} \\
\colhead{(1)} &
\colhead{(2)} &
\colhead{(3)} 
}
\startdata
A & 1.8-m Perkins Telescope + Ohio State CCD Spectrograph 	& 57 \\
F & 1.5-m Mt.\ Hopkins Telescope + CCD Spectrograph       	& 258 \\
GH& 2.1-m Guillermo Haro Observatory Telescope + B\&C
	Spectrograph						& 22 \\
H & 3.0-m Lick Shane Telescope + Kast Spectrograph             	& 28 
\\
K & 2.4-m MDM Observatory Hiltner Telescope +
	MODSPEC Spectrograph 					& 6 \\
L & 6-m Special Astrophys.\ Obs.\ Telescope + CCD Spectrographs & 19 
\\
L1 & 1-m Special Astrophys.\ Obs.\ Telescope + CCD Spectrographs & 32 
\\
M & 3.5-m and 2.2-m Calar Alto Telescopes + CCD Spectrograph    & 7 \\
R & 1.5-m Loiano Telescope + CCD Spectrograph                   & 1 \\
W & 2.6-m Shajn Telescope + CCD Spectrograph              	& 113 \\
\hline
  & {\bf TOTAL} & {\bf 543} \\
\enddata

\end{deluxetable}

%
%
\begin{deluxetable}{lll}
\tablewidth{0pt}
\tablecaption{Flux Scale Factors for Optical Spectra}
\tablehead{
\colhead{Data} &
\colhead{Point-Source} &
\colhead{Extended Source } \\
\colhead{Set} &
\colhead{Scale Factor} &
\colhead{Correction $G$} \\
\colhead{ } &
\colhead{$\varphi$} &
\colhead{($10^{-15}$ ergs s$^{-1}$\,cm$^{-2}$\,\AA$^{-1}$)} \\
\colhead{(1)} &
\colhead{(2)} & 
\colhead{(3)} 
}
\startdata
A 		& $1.000$ & $0.000$ \\
F (Year 8) 	& $0.941 \pm 0.035$ & $-1.730 \pm 0.050 $ \\
F (Years 9--13) & $1.039 \pm 0.042$ & $-0.532 \pm 0.570 $ \\
GH 		& $0.985 \pm 0.032$ & $-0.329 \pm 0.555 $ \\
H 		& $1.013 \pm 0.044$ & $-0.418 \pm 0.404 $ \\
K 		& $1.010 \pm 0.050$ & $0.234 \pm 0.448 $ \\
L 		& $1.003 \pm 0.056$ & $-0.874 \pm 0.645 $ \\
L1 (Year 10)	& $1.113 \pm 0.045$ & $4.088 \pm 0.396 $ \\
L1 (Year 11)	& $1.011 \pm 0.024$ & $1.500 \pm 0.931 $ \\
L1 (Year 12)	& $1.054 \pm 0.064$ & $2.556 \pm 0.844 $ \\
L1 (Year 13)	& $1.011 \pm 0.094$ & $3.090 \pm 0.954 $ \\
M 		& $0.990 \pm 0.014$ & $-1.353 \pm 0.380 $ \\
R 		& $0.863 \pm 0.025$ & $-1.900 \pm 1.065 $ \\
W (Year 8) 	& $0.932 \pm 0.031$ & $-1.319 \pm 0.437 $ \\
W (Years 9--13)	& $1.009 \pm 0.022$ & $-0.422 \pm 0.244 $ \\
\enddata
\end{deluxetable}

%
%
\begin{deluxetable}{llc}
\tablewidth{0pt}
\tablecaption{Sources of Photometric Observations, Years 9--13}
\tablehead{
\colhead{Data} &
\colhead{ } &
\colhead{Number of} \\
\colhead{Set} &
\colhead{Telescope and Instrument} &
\colhead{Observations} \\
\colhead{(1)} &
\colhead{(2)} &
\colhead{(3)} 
}
\startdata
A & 0.6-m Telescope, Crimean Laboratory of the Sternberg & 91 \\
  & Astronomical Institute +  pulse-counting& \\
  & UBV photometer, $14''\!.3$ aperture.    & \\
B & 1.25-m Telescope, Crimean Astrophysical Observatory & 12 \\
  & + pulse-counting Photometer-Polarimeter, $15''$ aperture& \\
C & 1.56-m Telescope, Shanghai Astronomical Observatory & 7 \\
  & + CCD, $10''$ aperture. & \\
D & 0.48-m and 0.6m Telescopes of Ulugbek Astronomical & 69 \\
  & Institute + pulse-counting UBV photometer, $14''\!.3$ aperture. & 
\\
E & 1-m and 0.6-m Telescopes of the Special Astrophysical & 67 \\
  & Observatory + CCD, $10''$ aperture. & \\
\hline
 & {\bf TOTAL} & {\bf 246} \\
\enddata

\end{deluxetable}

%
%
\begin{deluxetable}{ccc}
\tablewidth{0pt}
\tablecaption{Optical Continuum and H$\beta$ 
Light Curves\tablenotemark{a}}
\tablehead{
\colhead{Julian Date} &
\colhead{$F_{\lambda}$(5100\,\AA)}  &
\colhead{$F$(H$\beta$)}  \\
\colhead{($-2400000$)} &
\colhead{($10^{-15}$ ergs s$^{-1}$\,cm$^{-2}$\,\AA$^{-1}$)}  &
\colhead{($10^{-13}$ ergs s$^{-1}$\,cm$^{-2}$)}  \\
\colhead{(1)} &
\colhead{(2)} &
\colhead{(3)} 
}
\startdata
 50044.05 & $  11.52 \pm   0.23 $ & $   9.56 \pm   0.19 $ \\
 50048.60 & $  10.97 \pm   0.49 $ & $   8.78 \pm   0.23 $ \\
 50052.08 & $  10.77 \pm   0.27 $ & $   8.96 \pm   0.31 $ \\
 50053.01 & $  10.33 \pm   0.21 $ & $   9.02 \pm   0.18 $ \\
 50061.02 & $   8.56 \pm   0.17 $ & $   9.19 \pm   0.18 $ \\
 50064.60 & $   8.20 \pm   0.37 $ & $   8.33 \pm   0.22 $ \\
 50069.09 & $   8.18 \pm   0.20 $ & $   8.22 \pm   0.29 $ \\
 50074.00 & $   8.18 \pm   0.33 $ & $   7.27 \pm   0.20 $ \\
 50075.00 & $   8.30 \pm   0.33 $ & $   7.04 \pm   0.20 $ \\
 50078.00 & $   8.15 \pm   0.33 $ & $   6.92 \pm   0.19 $ \\
\enddata
\tablenotetext{a}{\raggedright A complete version of Table 5 is available at
{\sf http://www.astronomy.ohio-state.edu/$\sim$agnwatch/}}
\end{deluxetable}

%
%
\begin{deluxetable}{lcccccc}
\tablewidth{0pt}
\tablecaption{Sampling Statistics for Optical Continuum}
\tablehead{
\colhead{ } &
\colhead{Number of} &
\multicolumn{2}{c}{Sampling Interval (days)} &
\colhead{Mean} \\
\colhead{Subset} &
\colhead{Epochs} &
\colhead{Average} &
\colhead{Median} &
\colhead{Flux\tablenotemark{a}} &
\colhead{$F_{\rm var}$} &
\colhead{$R_{\rm max}$} \\
\colhead{(1)} &
\colhead{(2)} & 
\colhead{(3)} & 
\colhead{(4)} & 
\colhead{(5)} & 
\colhead{(6)} & 
\colhead{(7)} 
}
\startdata
All data      &1530 & 3.1 & 1.0 & $ 9.73\pm2.44$ & 0.247&$ 3.57\pm0.18$ 
\\
Year 1 (1989) & 125 & 2.4 & 1.0 & $ 9.92\pm1.26$ & 0.117&$ 2.16\pm0.16$ 
\\
Year 2 (1990) & 94  & 3.4 & 2.0 & $ 7.25\pm1.00$ & 0.129&$ 1.82\pm0.09$ 
\\
Year 3 (1991) & 65  & 4.8 & 3.0 & $ 9.40\pm0.93$ & 0.090&$ 1.51\pm0.09$ 
\\
Year 4 (1992) & 83  & 3.4 & 2.0 & $ 6.72\pm1.17$ & 0.168&$ 2.04\pm0.10$ 
\\
Year 5 (1993) & 174 & 1.3 & 0.7 & $ 9.04\pm0.90$ & 0.092&$ 1.65\pm0.08$ 
\\
Year 6 (1994) & 135 & 2.4 & 1.0 & $ 9.76\pm1.10$ & 0.104&$ 1.76\pm0.12$ 
\\
Year 7 (1995) & 83  & 4.0 & 1.9 & $12.09\pm1.00$ & 0.079&$ 1.48\pm0.04$ 
\\
Year 8 (1996  &	144 & 2.3 & 1.0 & $10.56\pm1.64$ & 0.150&$ 1.86\pm0.11$ \\
Year 9 (1997) & 126 & 2.4 & 1.0 & $ 8.12\pm0.91$ & 0.105&$ 1.84\pm0.06$ 
\\
Year 10 (1998)& 175 & 1.8 & 0.6 & $13.47\pm1.45$ & 0.100&$ 1.59\pm0.09$ 
\\
Year 11 (1999)& 148 & 2.1 & 1.0 & $11.83\pm1.82$ & 0.149&$ 1.98\pm0.11$ 
\\
Year 12 (2000)&  94 & 2.9 & 1.0 & $ 6.98\pm1.20$ & 0.166&$ 2.40\pm0.12$ 
\\
Year 13 (2001)&  84 & 4.6 & 2.0 & $ 7.03\pm0.86$ & 0.112&$ 1.68\pm0.10$ 
\\
\enddata
\tablenotetext{a}{Units of $10^{-15}$ ergs s$^{-1}$\,cm$^{-2}$\,\AA$^{-
1}$.} 
\end{deluxetable}

%
%
\begin{deluxetable}{lcccccc}
\tablewidth{0pt}
\tablecaption{Sampling Statistics for H$\beta$ Emission Line}
\tablehead{
\colhead{ } &
\colhead{Number of} &
\multicolumn{2}{c}{Sampling Interval (days)} &
\colhead{Mean} \\
\colhead{Subset} &
\colhead{Epochs} &
\colhead{Average} &
\colhead{Median} &
\colhead{Flux\tablenotemark{a}} &
\colhead{$F_{\rm var}$} &
\colhead{$R_{\rm max}$} \\
\colhead{(1)} &
\colhead{(2)} & 
\colhead{(3)} & 
\colhead{(4)} & 
\colhead{(5)} & 
\colhead{(6)} & 
\colhead{(7)} 
}
\startdata
All data      &1248 & 3.7 & 1.1 & $ 7.82\pm1.71$ & 0.216&$ 4.60\pm0.34$ 
\\
Year 1 (1989) & 132 & 2.3 & 1.0 & $ 8.62\pm0.85$ & 0.091&$ 1.57\pm0.12$ 
\\
Year 2 (1990) & 94  & 3.4 & 2.0 & $ 5.98\pm1.17$ & 0.191&$ 2.30\pm0.12$ 
\\
Year 3 (1991) & 65  & 4.8 & 3.0 & $ 7.46\pm0.81$ & 0.093&$ 1.58\pm0.14$ 
\\
Year 4 (1992) & 83  & 3.4 & 2.0 & $ 4.96\pm1.44$ & 0.284&$ 3.03\pm0.30$ 
\\
Year 5 (1993) & 142 & 2.1 & 1.0 & $ 7.93\pm0.53$ & 0.057&$ 1.40\pm0.06$ 
\\
Year 6 (1994) & 128 & 2.6 & 1.0 & $ 7.58\pm0.94$ & 0.117&$ 1.57\pm0.07$ 
\\
Year 7 (1995) & 78  & 4.2 & 2.1 & $ 9.27\pm0.70$ & 0.071&$ 1.38\pm0.06$ 
\\
Year 8 (1996) &	144 & 2.3 & 1.0 & $ 7.95\pm0.87$& 0.106&$ 1.76\pm0.08$ \\
Year 9 (1997) &  95 & 3.1 & 1.1 & $ 7.41\pm0.95$& 0.125&$ 1.66\pm0.07$ 
\\
Year 10 (1998)& 119 & 2.6 & 1.0 & $10.27\pm1.03$& 0.097&$ 1.50\pm0.07$ 
\\
Year 11 (1999)&  86 & 3.7 & 1.0 & $ 9.34\pm0.61$& 0.058&$ 1.50\pm0.06$ 
\\
Year 12 (2000)&  37 & 7.6 & 3.0 & $ 6.27\pm1.22$& 0.192&$ 1.96\pm0.07$ 
\\
Year 13 (2001)&  45 & 6.7 & 3.2 & $ 5.26\pm1.12$& 0.208&$ 2.33\pm0.08$ 
\\
\enddata
\tablenotetext{a}{Units of $10^{-13}$ ergs s$^{-1}$\,cm$^{-2}$.} 
\end{deluxetable}

%
%
\begin{deluxetable}{lccc}
\tablewidth{0pt}
\tablecaption{Cross-Correlation Results}
\tablehead{
\colhead{ } &
\colhead{$\tau_{\rm cent}$} &
\colhead{$\tau_{\rm peak}$} &
\colhead{ }  \\
\colhead{Subset} &
\colhead{(days)} &
\colhead{(days)} &
\colhead{$r_{\rm max}$}  \\
\colhead{(1)} &
\colhead{(2)} & 
\colhead{(3)} & 
\colhead{(4)}  
}
\startdata
Year 1 (1989) & $19.73^{+2.03}_{-1.40}$& $21.5^{+3.0}_{-4.1}  $& 0.877  
\\
Year 2 (1990) & $19.34^{+1.86}_{-2.96}$& $18.5^{+2.0}_{-0.7}  $& 0.909  
\\
Year 3 (1991) & $16.35^{+3.75}_{-3.28}$& $17.5^{+2.7}_{-5.7}  $& 0.740  
\\
Year 4 (1992) & $11.37^{+2.30}_{-2.30}$& $13.7^{+0.6}_{-4.7}  $& 0.918  
\\
Year 5 (1993) & $13.61^{+1.41}_{-1.91}$& $13.2^{+1.9}_{-3.3}  $& 0.730  
\\
Year 6 (1994) & $15.49^{+2.26}_{-6.09}$				     
                            & $\phantom{1}8.7^{+8.5}_{-2.4}   $& 0.832  
\\
Year 7 (1995) & $21.43^{+2.31}_{-3.00}$& $22.9^{+5.1}_{-3.1}  $& 0.880  
\\
Year 8 (1996) & $16.56^{+1.78}_{-1.08}$& $15.3^{+1.4}_{-1.3}  $& 0.907  
\\
Year 9 (1997) & $17.49^{+2.12}_{-1.74}$& $17.7^{+3.4}_{-1.8}  $& 0.803  
\\
Year 10 (1998)& $26.44^{+4.67}_{-2.63}$& $31.7^{+0.5}_{-10.9}  $& 0.844  
\\
Year 11 (1999)& $25.75^{+4.97}_{-3.32}$& $34.4^{+1.0}_{-18.5} $& 0.846  
\\
Year 12 (2000)& $ 6.12^{+4.44}_{-4.08}$& $ 8.0^{+5.4}_{-5.2}  $& 0.874  
\\
Year 13 (2001)& $14.89^{+5.30}_{-6.87}$& $ 8.0^{+14.6}_{-1.4}  $& 0.793  
\\
\enddata
\end{deluxetable}
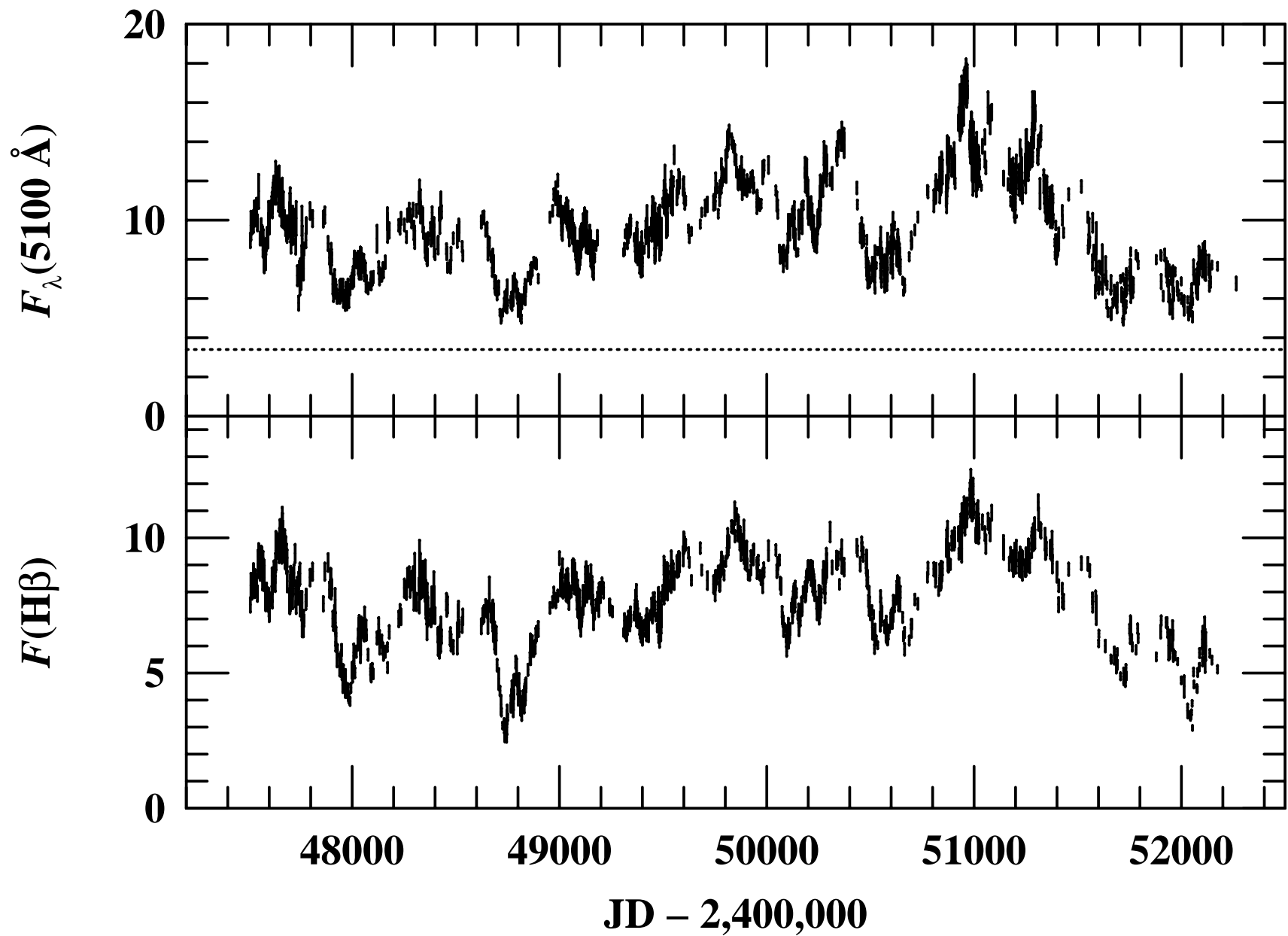
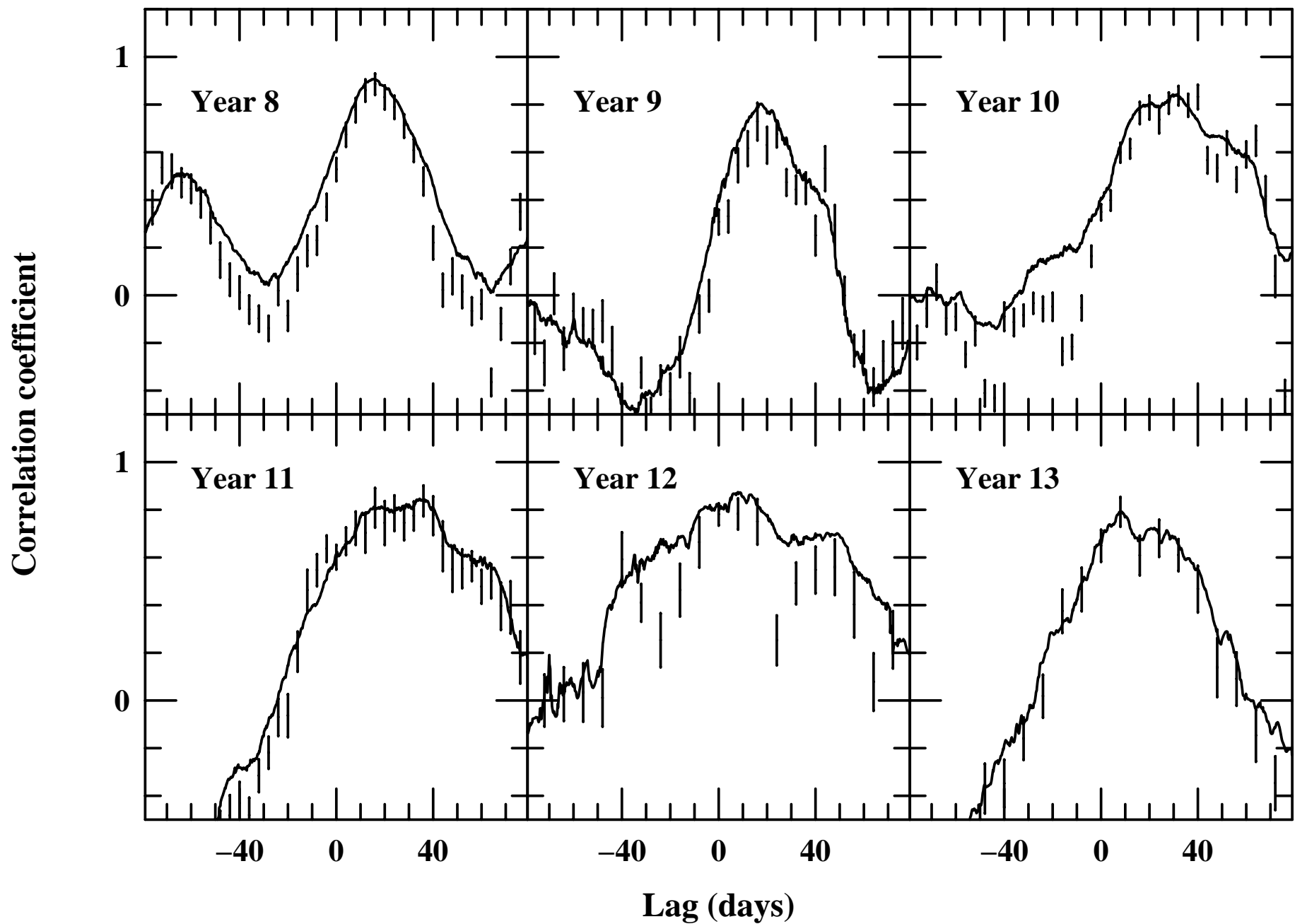
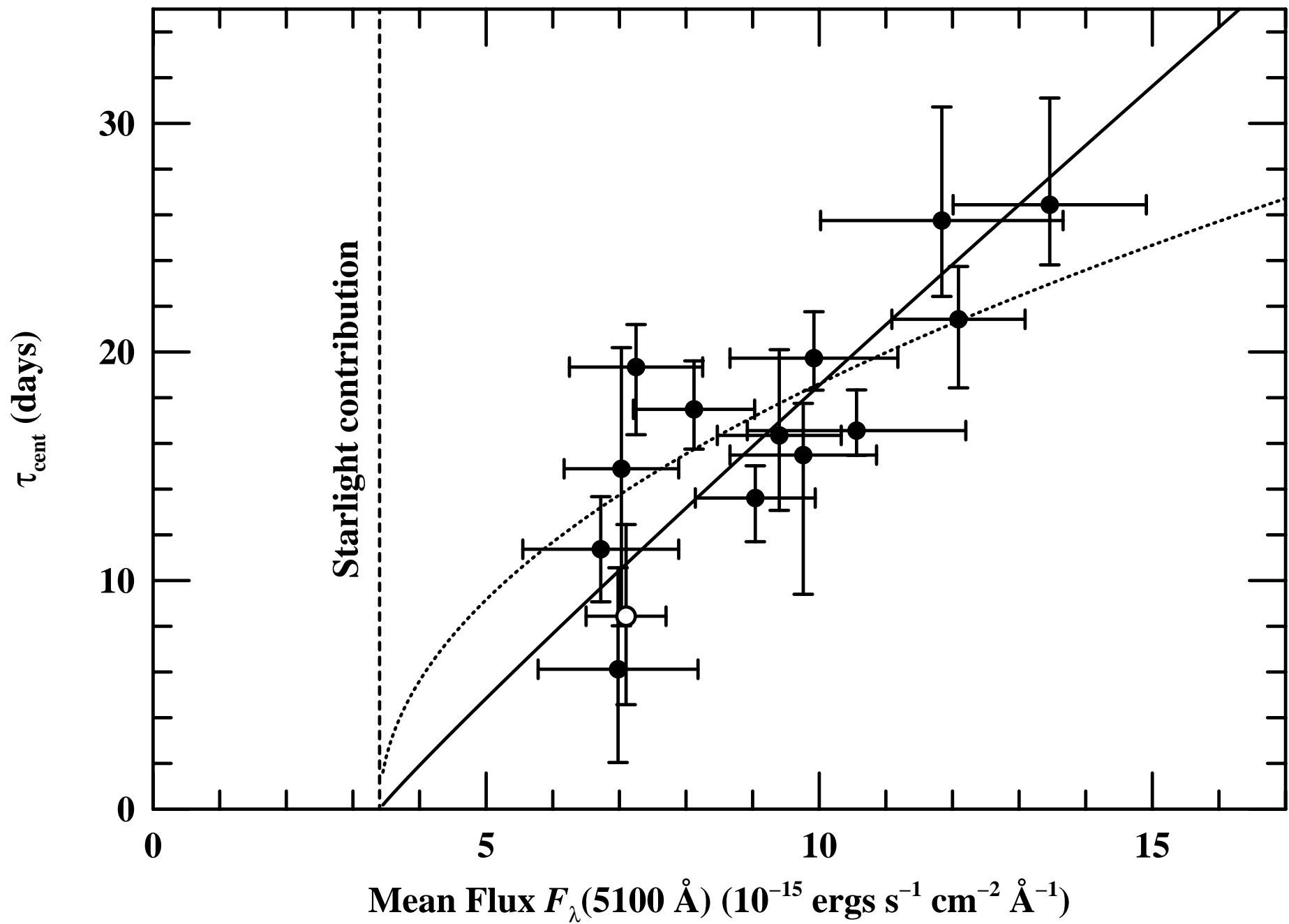
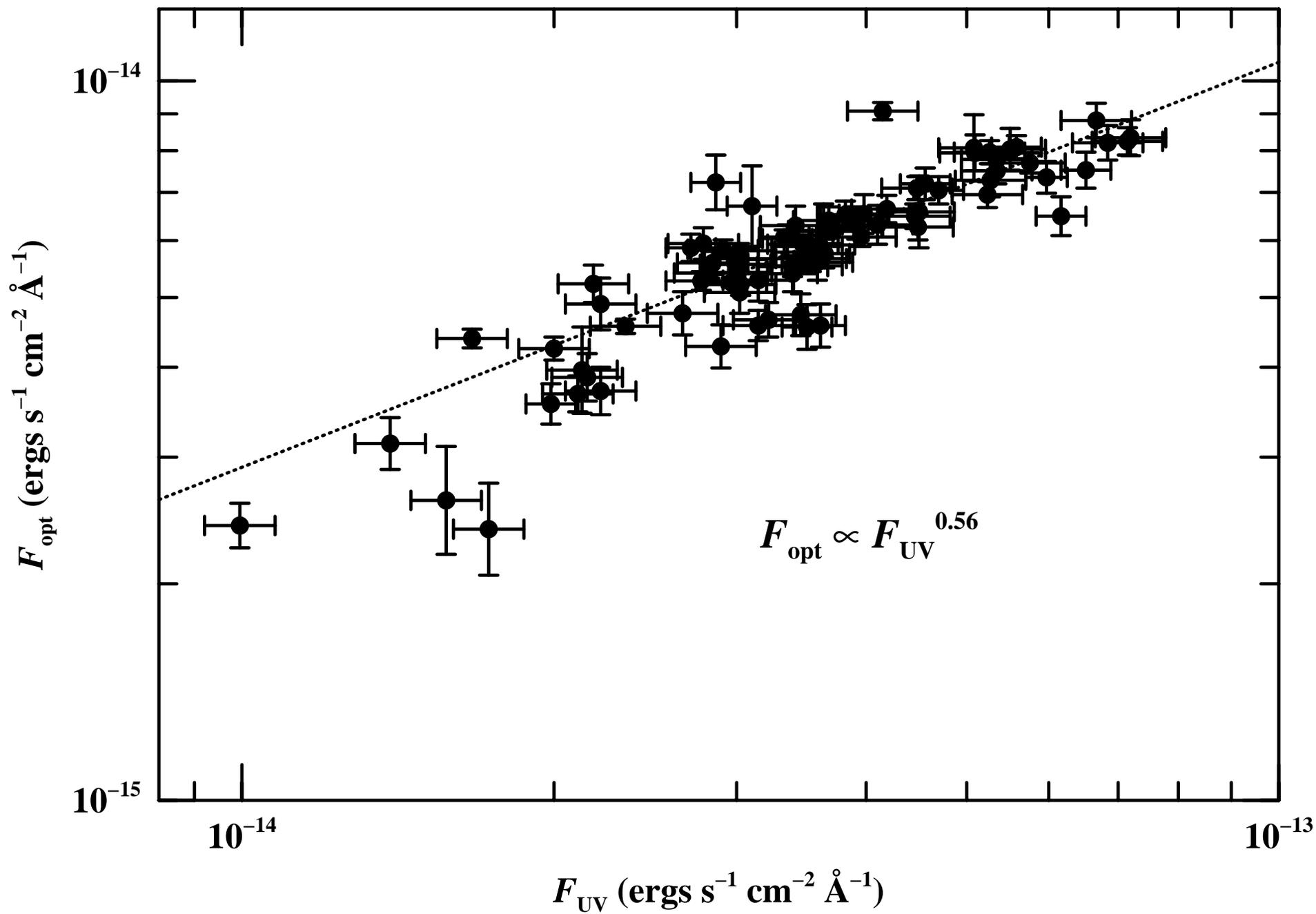
\end{document}